# Social stress drives the multi-wave dynamics of COVID-19 outbreaks

Innokentiy A. Kastalskiy[1,2,3,4,*], Evgeniya V. Pankratova[4,5], Evgeny M. Mirkes[4,6], Victor B. Kazantsev[1,3,4,7,8] & Alexander N. Gorban[1,4,6]

[1]Department of Neurotechnology, Lobachevsky University, 23 Gagarin Ave., 603022 Nizhny Novgorod, Russia

[2]Laboratory of Autowave Processes, Institute of Applied Physics of the Russian Academy of Sciences (IAP RAS), 46 Ulyanov St., 603950 Nizhny Novgorod, Russia

[3]Center for Neurotechnology and Machine Learning, Immanuel Kant Baltic Federal University, 14 Nevsky St., 236016 Kaliningrad, Russia

[4]Laboratory of Perspective Methods for Analysis of Multidimensional Data, Institute of Information Technology, Mathematics and Mechanics, Lobachevsky University, 23 Gagarin Ave., 603022 Nizhny Novgorod, Russia

[5]Department of Applied Mathematics, Institute of Information Technology, Mathematics and Mechanics, Lobachevsky University, 23 Gagarin Ave., 603022 Nizhny Novgorod, Russia

[6]Department of Mathematics, University of Leicester, University Rd, Leicester LE1 7RH, United Kingdom

[7]Neuroscience and Cognitive Technology Laboratory, Innopolis University, 1 Universitetskaya St., 420500 Innopolis, Russia

[8]Laboratory of Neuromodeling, Samara State Medical University, 18 Gagarin St., 443079 Samara, Russia

*email: kastalskiy@neuro.nnov.ru

## Abstract

The dynamics of epidemics depend on how people's behavior changes during an outbreak. At the beginning of the epidemic, people do not know about the virus, then, after the outbreak of epidemics and alarm, they begin to comply with the restrictions and the spreading of epidemics may decline. Over time, some people get tired/frustrated by the restrictions and stop following them (exhaustion), especially if the number of new cases drops down. After resting for a while, they can follow the restrictions again. But during this pause the second wave can come and become even stronger then the first one. Studies based on SIR models do not predict the observed quick exit from the first wave of epidemics. Social dynamics should be considered. The appearance of the second wave also depends on social factors. Many generalizations of the SIR model have been developed that take into account the weakening of immunity over time, the evolution of the virus, vaccination and other medical and biological details. However, these more sophisticated models do not explain the apparent differences in outbreak profiles between countries with different intrinsic socio-cultural features. In our work, a system of models of the COVID-19 pandemic is proposed, combining the dynamics of social stress with classical epidemic models. Social stress is described by the tools of sociophysics. The combination of a dynamic SIR-type model with the classical triad of stages of the general adaptation syndrome, alarm-resistance-exhaustion, makes it possible to describe with high accuracy the available statistical data for 13 countries. The sets of kinetic constants corresponding to optimal fit of model to data were found. These constants characterize the ability of society to mobilize efforts against epidemics and maintain this concentration over time, and can further help in the development of management strategies specific to a particular society.

## Introduction

In December 2019, a growing number of severe cases of viral pneumonia were reported in Wuhan city, Hubei Province, People's Republic of China. The virus spread around the world in about three months. Therefore, on March 11, 2020, the World Health Organization declared a global pandemic. A novel type of coronavirus, SARS-CoV-2, led to tremendous changes in economy, education and our daily life.

From the beginning of the pandemic, many mathematical models have been proposed to simulate the virus spread and possible ways of its control[1-6]. The SIR-type models and their modifications are widely used to describe qualitatively the infection outbreak in homogeneously mixed populations. In particular, various SEIR-type[2,7,8] (susceptible-exposed-infected-removed), SEIQR-type[4] (additional quarantined), SEIR models with delays[5,9,10], SIR-type time-dependent[11] and compartmental models[12], or even a SEIIHURD model[13] with the proportion of symptomatic needing hospitalization and those in critical conditions requiring intensive care admission, and many others accounting different particular factors have been proposed. Various complex multidimensional models of social contacts were considered. For example, one of the heterogeneous SAIR-models[14] (susceptible-asymptomatic-infected-removed) took into account the different numbers of contacts between individuals in the network with various degrees of the network nodes.

The emergence of the second and consequent peaks in the dynamics of epidemics and effective ways to control them are of particular interest for the analysis of the latest data on pandemics. The first epidemic wave decays more rapidly than the SIR-type models prescribe, and then a second wave emerges that is impossible in SIR models. In a recent work[15], an eight-dimensional SEIR-type model was proposed to analyze the ways to reduce the rate of SARS-CoV-2 virus transmission. Based on its analysis the double-peak behavior caused by removing restrictions was demonstrated[3]. It was also shown that the second peak could be much higher than the first one.

Other studies[16,17] focused directly on the dynamics of the second wave. They performed a comparative analysis of trajectories for different regions and identified the countries that most significantly reduced the mortality rate.

Various stress factors can significantly modify social dynamics[18]. In the modern world, TV, the Internet and social networks provide instant transmission of information. Information spreading in times of crisis affects human behavior and may create "infodemics"[19]. Changes in people's daily lives, either due to their own fears or due to forced administrative restrictions, lead to a significant change in the effective reproduction number of a viral disease. In a recent research[6], the entire population was divided into the groups where the persons either obey or ignore social distancing rules. This assumption made it possible to determine the size of these groups, depending on social factors. Oscillatory patterns of repeated indulgences (when the situation seems to be improving) and tightening of restrictions (when the spread of infection becomes high) were found.

General adaptation syndrome (GAS) was discovered by Selye (1936)[20]. He found that "a typical syndrome appears, the symptoms of which are independent of the nature of the damaging agent". Later on, this concept was integrated with Cannon's *fight-or-flight response*[21], and the general concept of stress was developed with a wide range of applications in physiology, psychology and beyond[22-24]. We will use these results as prototypes for our models.

GAS consists of three phases: a mobilization phase, *alarm*, a *resistance* phase, and an *exhaustion* phase[25]. To simulate the population dynamics within the conditions of aggressive COVID-19 spreading, the susceptible group of the persons can be divided into three subgroups of the individuals according to their *behavior modes*. The models of behavior in which people are seen as behaving in different ways (modes) at different times and in different contexts is widely used in the game theory and analysis of economic and social behavior[26]. According to the concept of GAS and stress, we consider three modes: (i) *ignorant* mode (persons living without any restrictions as it was before the pandemic), (ii) *resistance* mode (conscious persons practicing the social *distancing* rules to avoid the appeared danger) and (iii) *exhaustion* mode (the depletion of the person's resources leading to reduction of following social distancing rules). In the first approximation, we exclude the *alarm* mode that may be considered as a short delay before activation of resistant behavior. In more detailed models, alarm mode can be considered separately, but the more phases we consider, the more unknown constants we have to identify, and the simplest model should be tested first. We consider a loop of transitions between modes:

$$ignorant \to resistance \to exhaustion \to ignorant.$$

This loop should be closed: after some time persons return from the exhaustion stage into initial mode and are susceptible to the alarm signals again. The probability of infection is much less in the resistant mode than in the ignorant or exhaustion ones.

The inclusion of social stress in the SIR type model gives rise to a new model, which we call $SIR_{SS}$. It describes the multi-peak dynamics in the COVID-19 outbreaks. The model parameters have been successfully fitted to best match the statistical observations of epidemics in different countries of the world.

The parameters of social dynamics found for different countries demonstrate the differences between their social reactions to pandemic stress. These parameters are: $a$ – morbidity rate, $K_2$ – exhaustion rate, $q$ – stress response rate, $I_0$ – initial fraction of infected population. The qualitative picture meets the general expectations. Particularly instructive are the quantitative differences between countries. The fitted models can be used in the development of management strategies specific to a particular society. The modeling method can be directly transferred and applied to regions, cities and sufficiently large subpopulations.

**Results**

**Comparative study concept and main results.** We used simulations to fit the total cumulative cases (TCC) of COVID-19 observations in China and 12 other countries with a large number of coronavirus cases: Brazil, Colombia, France, Germany, India, Iran, Israel, Italy, Russia, Spain, UK, and USA. We took the duration of the simulation, which fully covers the first wave of the epidemic, but not too long so that it is not affected by seasonal fluctuations in immunity. An interval of 200 days was chosen as the priority period for simulations. We used the calculation of the $R^2$ coefficient as a goodness-of-fit measure to perform global optimization in the parameter space ($a$, $K_2$, $q$, $I_0$). Finally, we have identified the precise values of these parameters, which provide the best fit between the theoretical curve of cumulative cases and the observed one. The results are shown in **Table 1**.

**Table 1 Key epidemiological parameters used in simulation to explain the COVID-19 observations**

| Country or territory | Population | Number of days modeled | Parameters | | | | $R^2$ |
| :---: | :---: | :---: | :---: | :---: | :---: | :---: | :---: |
| | | | $a$ | $K_2$ $\times 10^{-3}$ | $q$ $\times 10^3$ | $I_0$ $\times 10^{-6}$ | |
| China | 1,439,324 K | 200 | 0.2902 | 2.67 | 378370 | 1.00 | 0.99460 |
| Brazil | 212,559 K | 300 | 0.1371 | 15.36 | 3.21 | 36.6 | 0.99911 |
| Colombia | 50,883 K | 300 | 0.1451 | 20.84 | 5.84 | 5.72 | 0.99971 |
| France | 68,148 K | 200 | 0.3138 | 5.43 | 213.5 | 0.36* | 0.98727 |
| Germany | 83,784 K | 200 | 0.2190 | 6.06 | 292.7 | 18.0 | 0.99898 |
| India | 1,380,004 K | 400 | 0.1346 | 6.76 | 27.0 | 2.64 | 0.99730 |
| Iran | 83,993 K | 100 | 0.2096 | 20.36 | 1322 | 11.5 | 0.99910 |
| Israel | 8,656 K | 100 | 0.2473 | 7.85 | 397.8 | 34.8 | 0.99311 |
| Italy | 60,462 K | 200 | 0.1929 | 5.22 | 87.1 | 38.4 | 0.99713 |
| Russia | 145,934 K | 200 | 0.1495 | 13.45 | 47.5 | 35.9 | 0.99878 |
| Spain | 46,755 K | 200 | 0.3770 | 4.56 | 75.4 | 1.17* | 0.98844 |
| UK | 67,886 K | 200 | 0.1798 | 7.27 | 83.7 | 28.5 | 0.99828 |
| USA | 331,003 K | 150 | 0.2065 | 19.23 | 85.2 | 19.2 | 0.99899 |
| **Median** | **83,784 K** | **200** | **0.2065** | **7.27** | **85.2** | **18.0** | **0.99828** |

*Unusually small $I_0$ for France and Spain may be caused by numerous data corrections in April'20-May'20

**Fig. 1** presents the results of modeling.

The notation used in the figures: TCC – the total number of cumulative cases (observed); DNC – daily new cases (observed); CC(t) – the predicted values of the cumulative cases; CC'(t) – the predicted daily new cases; the normalized values of CC and CC' are the fractions of CC and CC' in the population; S – the susceptible fraction of the population; $S = S_{ign} + S_{res} + S_{exh}$ for three modes of behavior: ignorant, resistance, and exhaustion; I – the infected fraction of the population; R – the removed fraction of the population.

In China, the resistance index $q$ is unprecedentedly higher than in other densely populated countries. This immediately indicates that the highest discipline and state control that we observed in China allowed this country to avoid the second wave of the coronavirus outbreak (**Fig. 1**).

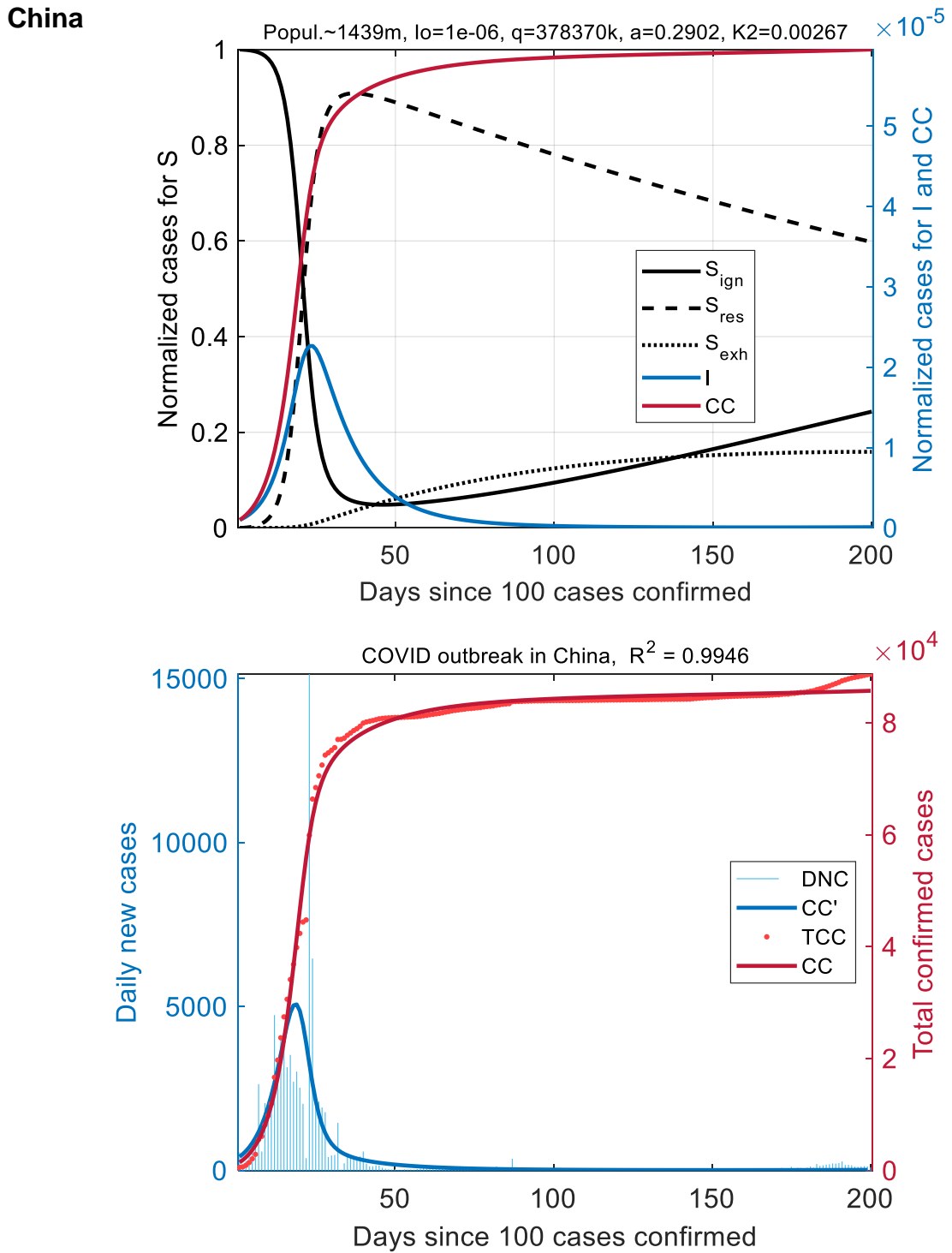

**Fig. 1 Coronavirus outbreak in China.** The dynamics of the SIR$_{SS}$ model is shown in the **top panel**: $S_{ign}$, $S_{res}$, $S_{exh}$, I, and CC = I + R. Fitted COVID data in absolute values (total confirmed and daily new cases: TCC and DNC) are displayed at the **bottom**. The beginning of the infection spread is characterized by a single wave, followed by a plateau on CC(t). Fractions $S_{ign}$ and $S_{res}$ quickly interchange with an increase in I(t), and $S_{exh}$ gradually increases to a relatively low level of ~15%.

**The first COVID-19 outbreaks in Europe.** The general sample of 12 countries was divided into four groups of three countries each. First, we analyzed the large European countries, where the coronavirus outbreak occurred at the beginning of the pandemic: Italy, Germany, and United Kingdom (**Fig. 2**). An interval of 200 days was also chosen for the simulations.

### Italy

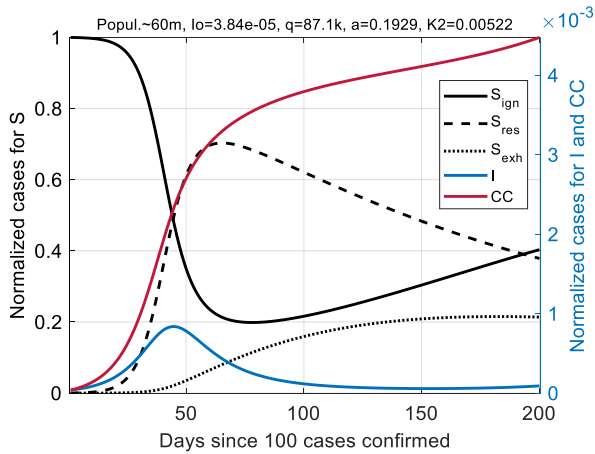
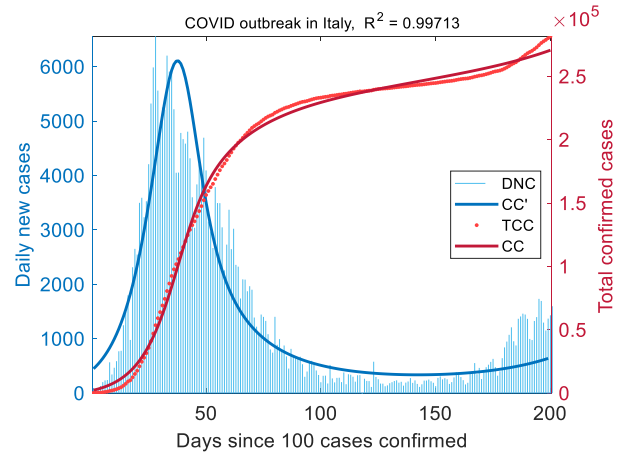

### Germany

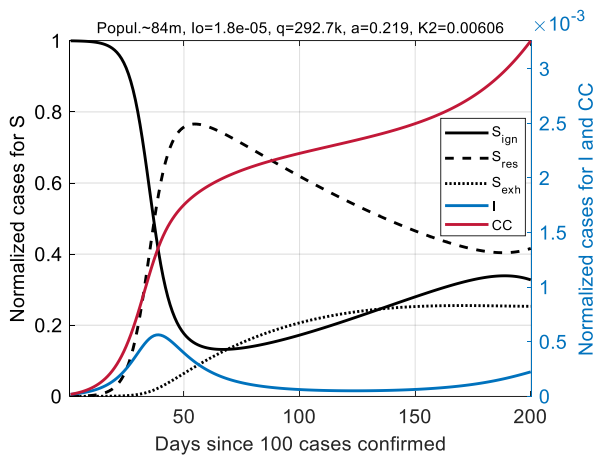
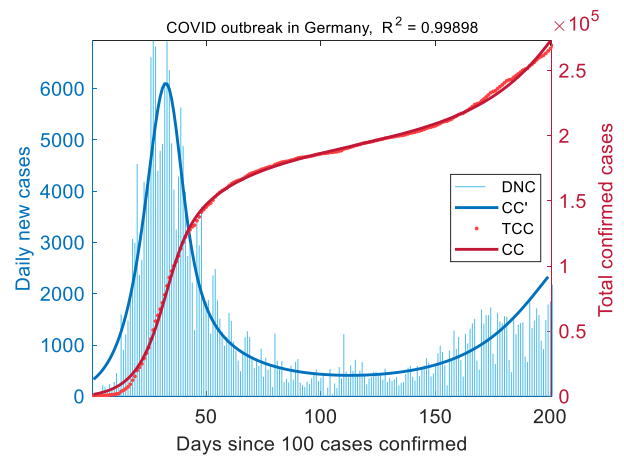

### United Kingdom

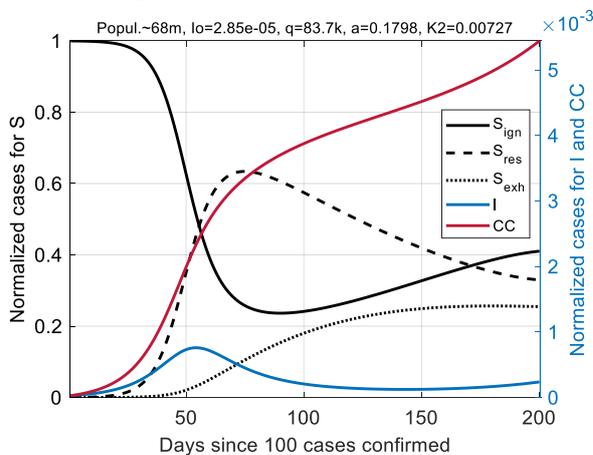
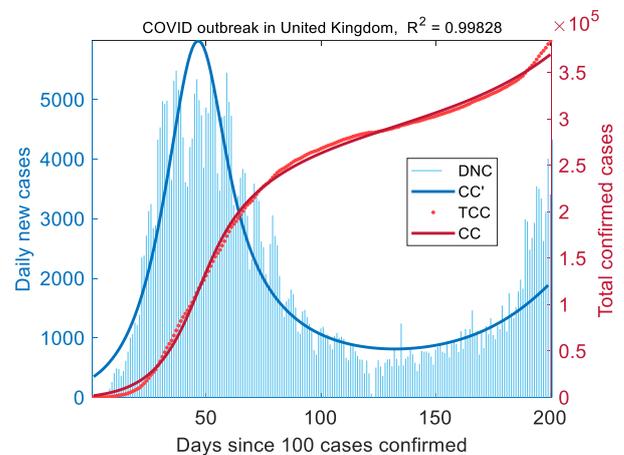

**Fig. 2 Earliest coronavirus outbreaks in Europe.** Results for Italy (**top panel**), Germany (**center**), and United Kingdom (**bottom**). The dynamics of the epidemic spread are qualitatively similar and are characterized by a profile containing the full first wave and the beginning of the second wave.

The results demonstrate an almost complete match of the cumulative cases, CC(t), for the SIR$_{SS}$ model data and the COVID-19 statistics, TCC, ($R^2 > 0.997$). The dynamics of the epidemic spread were accompanied by antiphase oscillations of $S_{ign}(t)$ and $S_{res}(t)$ with sustained leakage into the exhausted fraction $S_{exh}(t)$. In **Fig. 2**, the number of $S_{exh}$ reaches the saturation level after the end of the first epidemic wave. Curves for the infected state, I(t) and DNC(t) = TCC'(t), show the two-wave dynamics. The SIR$_{SS}$ model demonstrates the effect of reduction in morbidity at the peak of the first wave due to the rapid increase in the resistant fraction $S_{res}(t)$. Consequently, the I(t) dependence enters the second wave due to the accumulation of social stress and exhaustion caused by restrictions. This distinguishes our SIR$_{SS}$ model from the classical SIR, characterized by the depletion of potential carriers of the infection right after the first wave. For example, in the simplest SIR model with two parameters (*a*, *b*), only one epidemic wave I(t) can exist. The function CC(t) = I(t) + R(t) is a sigmoid with a saturation level of 32% of the population at parameters (0.12, 0.1) and 80% of the population at (0.2, 0.1). Observations of COVID-19 epidemics show that the first wave of the disease does not reach such high values and stops much earlier.

**Countries with a high resistance index.** Next, we analyzed the infection data of COVID-19 for countries where the coronavirus outbreak evolved dramatically like in China: it was relatively short-lived, and after it, the morbidity rate DNC(t) rapidly dropped to a low level. For this purpose data about France, Israel, and Spain were selected and analyzed (**Fig. 3**). Several trends were revealed from the results in **Table 1**. The parameter *q*, which is responsible for the mobilization response, for France and Israel is several times higher than the median value $q^* = 85.2$. At the same time, in France and Spain, the infection rate, *a*, takes on abnormally high values: 0.3138 and 0.3770 relative to the median, $a^* = 0.2065$. Thus, the tendency of society to trigger into the *alarm* state, combined with a high infection rate, leads to highly unified society's response[24]. Note, that at the peak the fraction of actively resistant people $S_{res}$ can rise up to 90% of total population.

## France

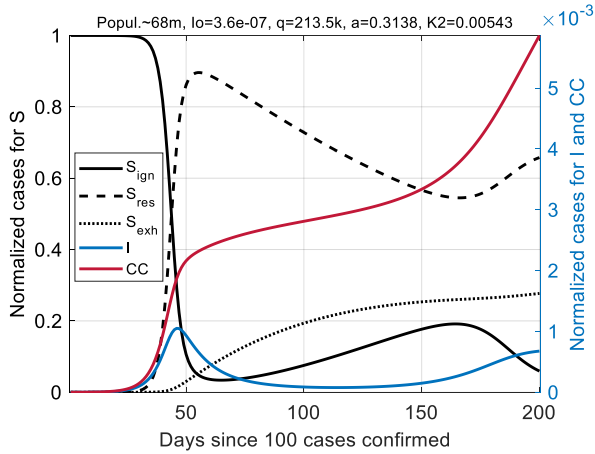
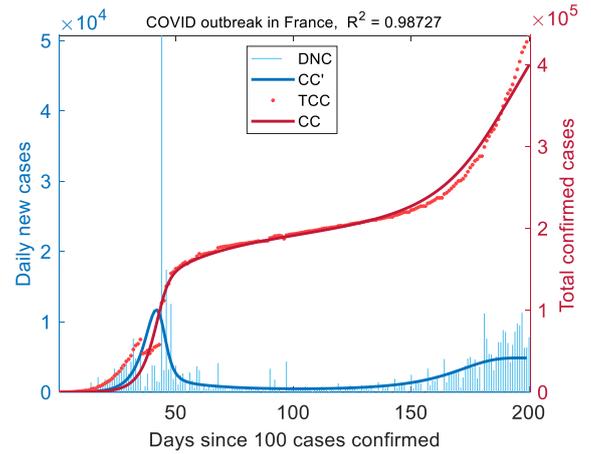

## Israel

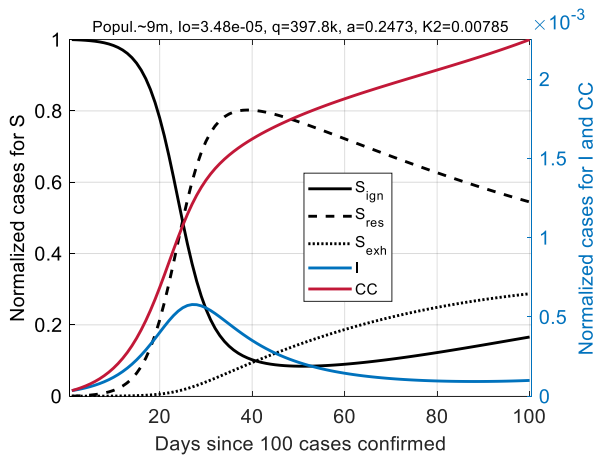
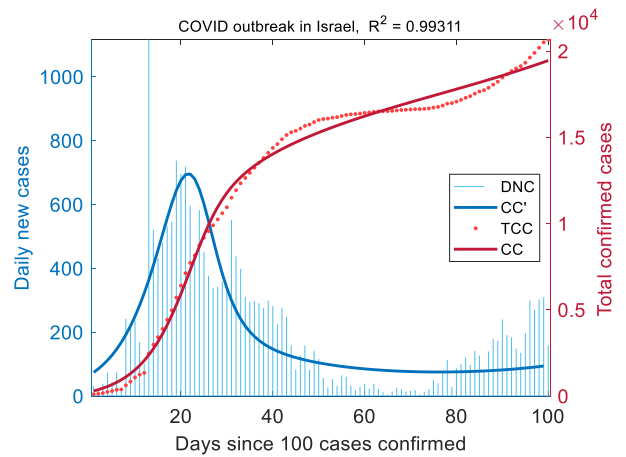

## Spain

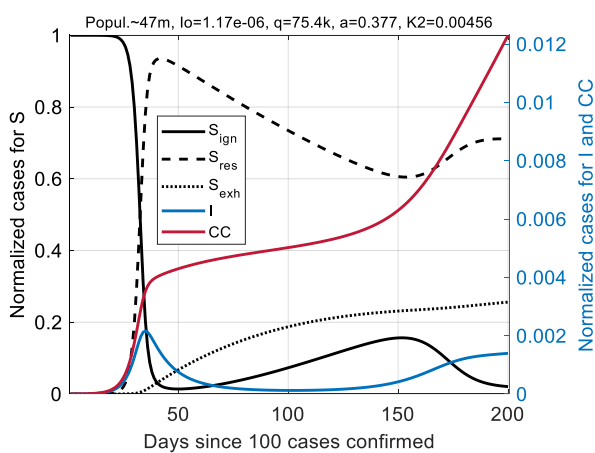
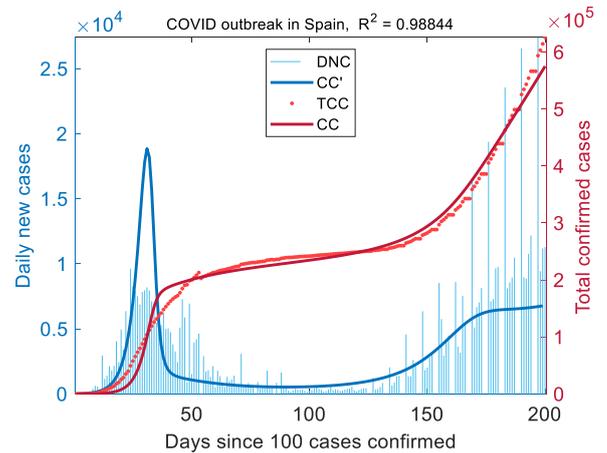

**Fig. 3 Coronavirus outbreaks in countries with a high resistance index.** Results for France (**top panel**), Israel (**center**), and Spain (**bottom**). The outbreaks dynamics are characterized by a rapid increase in the $S_{res}$ fraction during the peak of the first wave and the formation of protracted plateau on CC(t) with low DNC indices. To simulate the spread of COVID in Israel (population is less than 10 million), we take an interval of 100 days. Numerous data corrections in France and Spain for the period April'20-May'20 lead to the minor (practically negligible) discrepancies between the CC and TCC profiles.

**Countries with a low resistance index.** On the contrary, when the parameter $q$ takes small values, the behavioral reaction of the population to the growth of I(t) is weakly expressed. This case corresponds to the third group of countries selected for analysis: Brazil, India, and Russia (**Fig. 4**). At $q \ll q^*$, the similarity of the SIR$_{SS}$ model with the classical SIR is high. Combined with the low infection rate, it leads to the "lengthening" of the first epidemic wave in the timeline. The fraction of resistant population, $S_{res}(t)$, over the entire interval of the outbreak for these countries is less than 40%.

The $K_2$ parameter for India, which determines the predisposition to trigger into exhaustion mode, is close to this parameter for all five analyzed European countries. Nevertheless, in India this did not prevent $S_{exh}(t)$ from evolving at the lowest level among the countries of the general sample, not exceeding 20%, due to the constantly low level of the fraction $S_{res}$ (low resistance − low exhaustion). This parameter for Russia ($13.45\times10^{-3}$) and Brazil ($15.36\times10^{-3}$) was significantly higher than the median $K_2^* = 7.27\times10^{-3}$. However, the saturation level of the exhaustion fraction for Russia and Brazil was comparable to Western European countries with minimal $K_2$: $S_{exh}^* \sim 30\%$.

The distinctive feature of this group was the lower morbidity rate $a$, which was in the top-3 of the lowest values of $a$. It required much longer time intervals for simulation (up to 400 days). The persistence of the high fraction $S_{ign}$, in contrast to the countries of the second group, can be a precondition for the imminent emergence of the second, more substantial epidemic wave without the rest period between waves. Compare the DNC(t) trajectory with a low plateau in **Fig. 3** and the trajectories without such a plateau in **Fig. 4**. This fact corresponds to the official COVID-19 statistics.

## Brazil

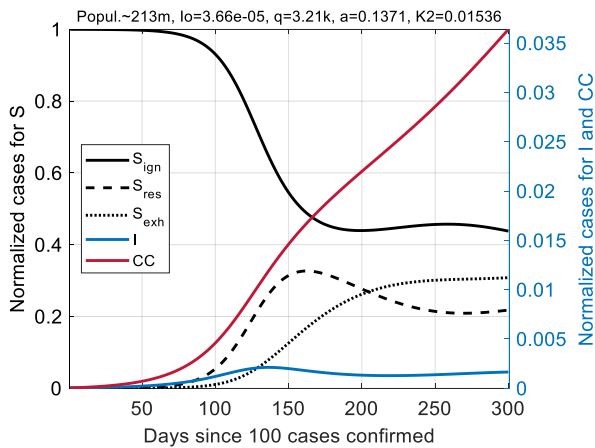 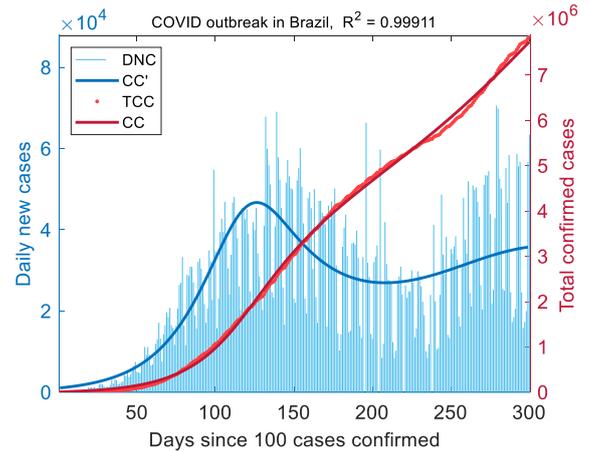

## India

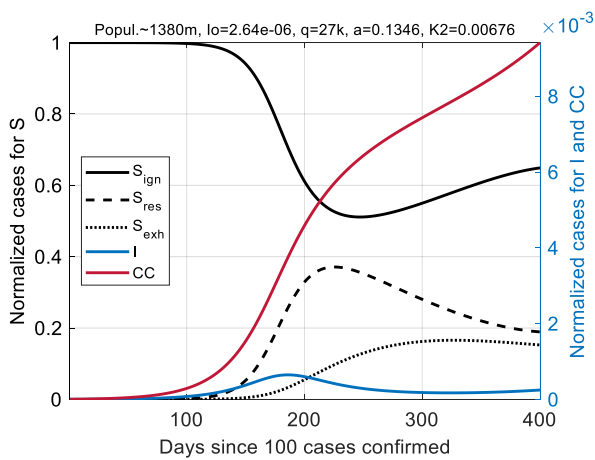 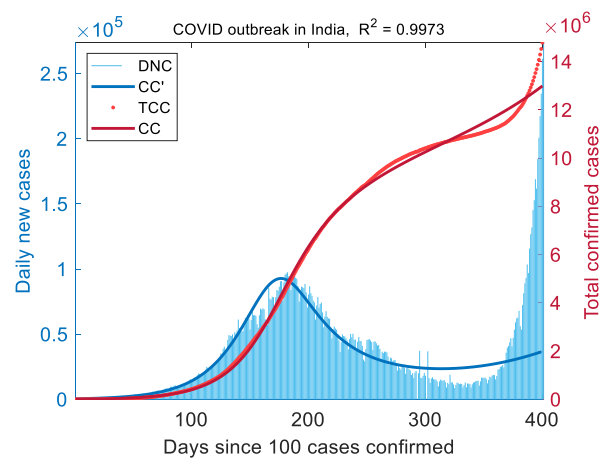

## Russia

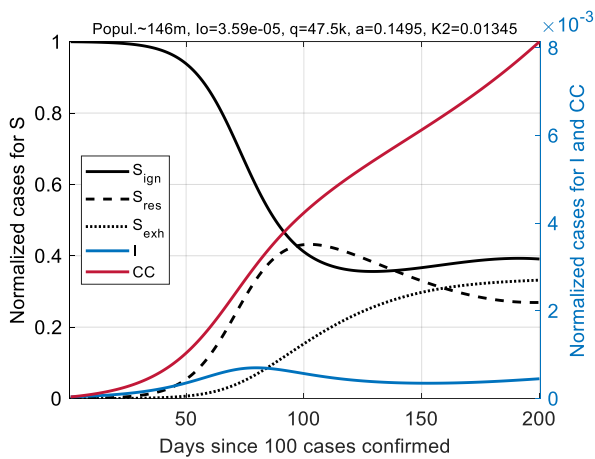 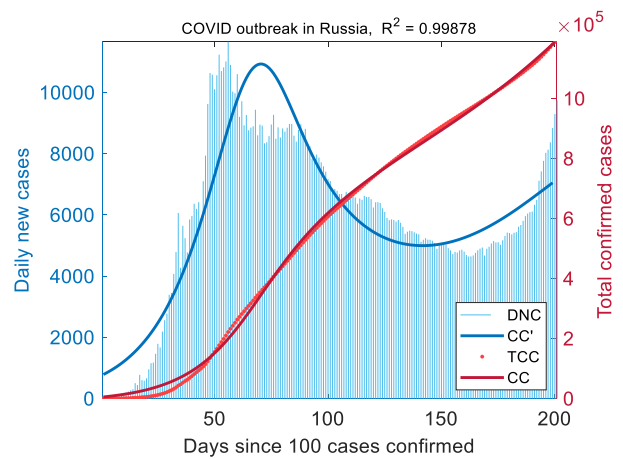

**Fig. 4 Coronavirus outbreaks in countries with a low resistance index.** Results for Brazil (**top panel**), India (**center**), and Russia (**bottom**). Trajectories are characterized by a low $S_{res}$ and a high $S_{ign}$ in comparison with the previous groups. The exhausted fraction of the population $S_{exh}$, on the contrary, is quantitatively comparable and evolves in approximately the same scenario. The end of the first wave I(t) smoothly turns into the second one. To simulate the spread of COVID-19 in Brazil and India, which have large populations, intervals of 300 and 400 days were taken, respectively.

**Countries with a high exhaustion rate.** Next, we analyzed COVID-19 data for the remaining three countries: Colombia, Iran, and USA (**Fig. 5**). This group was characterized by relatively high values of the exhausted population, $S_{exh}$ ~ 50%. As expected, the parameter $K_2$ ~ $20 \times 10^{-3}$ took extremely high values; this group is in the top-3 countries from the general sample in terms of the $K_2$ coefficient. Ordinary infection rates $a$ ~ 0.15-0.2 indicated the average growth rate of the cumulative cases number. Large scatter of values of $q$ illustrates the weak dependence of CC(t) trajectory on this parameter.

Similarly to the previous group, the formation of the second epidemic wave was also revealed. However, in this case, the forecast foreshadows a "huge" wave in which the DNC values may be many times greater than the first peak.

## Colombia

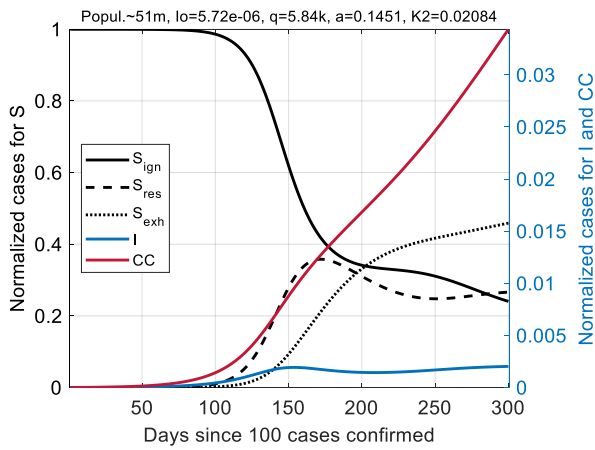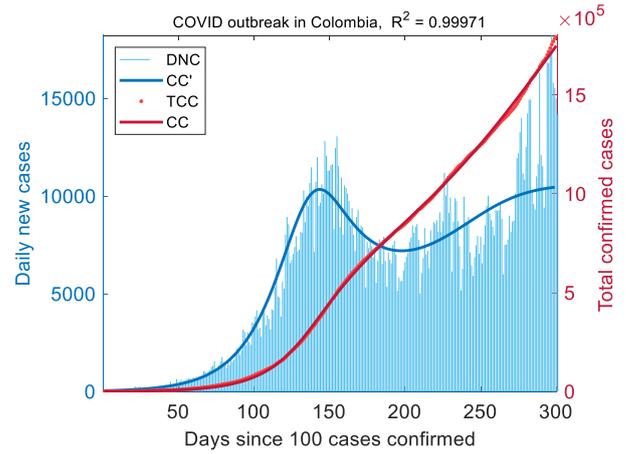

## Iran

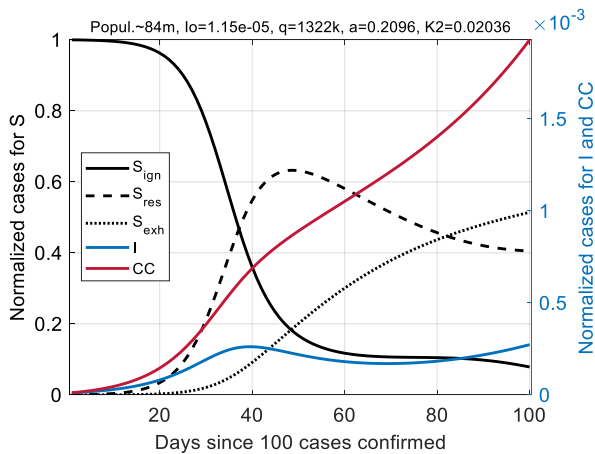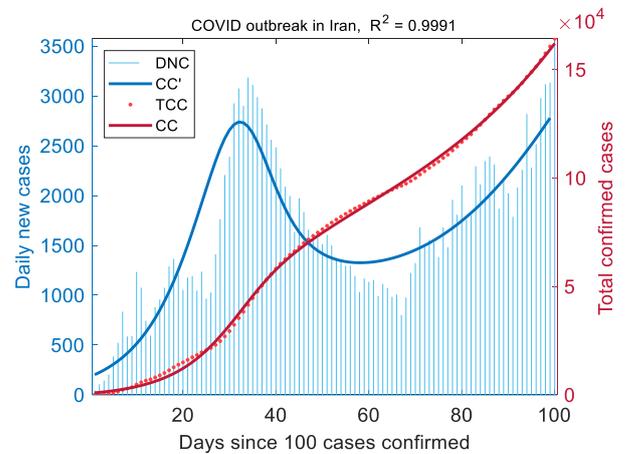

## United States

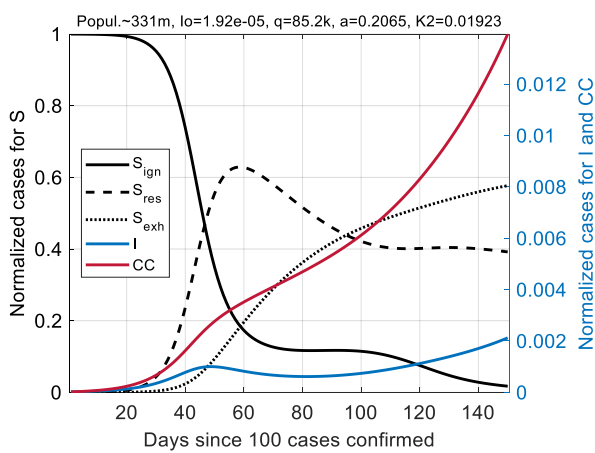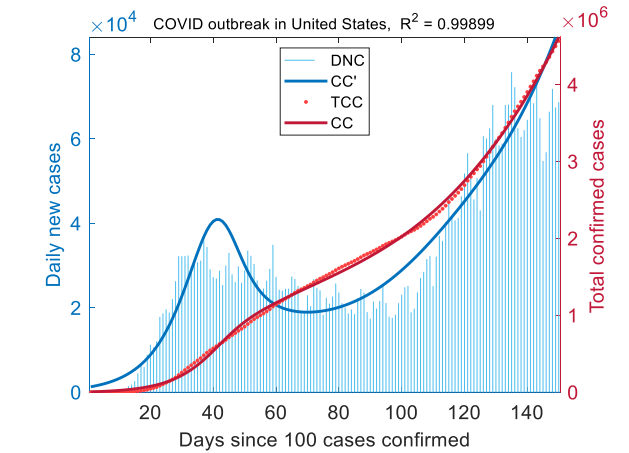

**Fig. 5 Coronavirus outbreaks in countries with extremely high exhaustion rate.** Results for Colombia (**top panel**), Iran (**center**), and United States (**bottom**). The main trend here is the rapid accumulation of exhausted people $S_{exh}$ (up to ~50% over a period of 50-150 days, if taken from the peak of the first wave). At the same time, the $S_{ign}$ fraction can remain quite small. The first wave of $I(t)$ rapidly turns into the second one without a significant decline relative to the peak values. The time intervals for the simulations were chosen empirically.

## Discussion

We demonstrated how the dynamics of the COVID-19 epidemics are mainly driven by the dynamics of social stress. The large difference between epidemics in different countries is caused by social differences and not by real epidemiological (biological) factors, at least until new local viral mutations appear.

The simple $SIR_{SS}$ model works well for 200-300 days of pandemics. We cannot expect that such a simple model will work much longer for various reasons. It has limitations and does not take into account many factors that arise during epidemics, for example:

- Improvement of medical and biological protection methods like new medicine and vaccination;
- Changing government decisions affecting the behavior of millions of people;
- Economic trends;
- Viral mutations.

All these changes require modification of the coefficients of the model and, more importantly, entail the introduction of more processes, such as dynamics of attitudes towards vaccination, social conflicts, and changes in economics.

It may seem rather surprising that the simple $SIR_{SS}$ model, which hardly claims to account for all the epidemiological details of COVID-19, was flexible enough to describe the response of society to virus intervention and subsequent adaptation. At the scale of countries the model was sufficient to describe different kinds of statistics of available data with high accuracy. However, we should note that only available official COVID-19 statistics were taken into account in this research. The $SIR_{SS}$ model is not intended to predict how many times the true fraction of the infected population exceeds the number of total confirmed cases.

Our model allows us to assess the development of the epidemic and strategies to combat it for countries with different social structures and cultural traditions. This possibility could be more important than making predictions. The difference between countries is very clearly represented in kinetic constants.

The model parameters were fitted to various empirical data using a global optimization algorithm (a kind of machine learning). The parameters of social stress response in different countries were estimated. Two of them vary significantly between countries and characterize them: $K_2$ (exhaustion rate or the inverse time of exhaustion) and $q$ (stress response or mobilization rate). The results (**Table 1**) can be summarized as follows.

- A very low $K_2$ value combined with an extremely high $q$ allows countries to avoid the second wave and remain with the classic single peak profile in the dynamics of the infected fraction, I(t).
- A very high $K_2$ will inevitably lead to a double-peaked wave, where the second can be particularly large.
- A low $K_2$ value, together with a high $q$ and a high $a$ (morbidity), provokes two split waves at I(t) with a protracted plateau at CC(t).

The choice of the simplest model of epidemics driven by social stress was almost unambiguous. There are many avenues for future research. Many questions should be answered about the dynamics of immunity, the evolution of viruses, the dynamics of opinion, the logic of government decisions, feedback from social stress to governments, the dynamics of the economy during pandemics, and many others. Each complex model will have many parameters. They need to be determined from the data and the reliability of each value can be questioned. The trade-off between the model complexity and the reliability of parameter determination is a special problem to be solved. The key issue is the combination of modeling social processes with the dynamics of immunity, viral evolution and economics.

There are also several particular technical problems that are important for COVID-19 epidemics data analysis and modeling. For example, this is a question about hidden cases: a significant proportion of infected people are not counted in the TCC. The proportion of latent cases is not constant and depends on the workload of health services (and its possible overload), the organization of testing, the intensity of epidemics and other problems. To estimate the proportion of latent cases and to correct the TCC, special work on data analysis and modeling is required, in order to fit CC(t) not to the registered, but to the corrected TCC.

The main result of the research program should be a system of models that will provide us with tools for quantifying various situations and playing out various scenarios *in silico* to develop anti-epidemic strategies specific to a particular society: country, region or social group.

## Methods

**SIR$_{SS}$ model features description.** SIR-type models mainly describe the following two types of transitions: $S + I \rightarrow 2I$ and $I \rightarrow R$. The first one defines close contact infection of a susceptible person with an infected person. The second one describes the relaxation process of recovery or death from the disease. The reaction rate obeys the law of mass action, which sociophysics borrowed from chemical kinetics. The autocatalytic form $S + I \rightarrow 2I$ means that this is a transition $S \rightarrow I$ with the transition rate proportional to the fraction of infected people I. In state R, a person is no longer contagious and does not return back to state S, therefore, for relatively short periods of time (for example, no more than a year), one can take total population as constant: $S(t) + I(t) + R(t) = 1$.

However, the basic SIR models do not take into account a person's reactions to the information flow and events taking place in society. In particular, important administrative decisions, which require stricter hygiene standards and social distancing, affect human behavior and disease transmission.

We introduce an enhancement of the SIR model that takes into account the impact of social stress factor[25]. According to the general adaptation syndrome (GAS) theory, there are three chronological stages of the stress response: *alarm*, *resistance*, and *exhaustion*. After some rest the exhausted person can return to the initial state and become susceptible to alarm. Given the fraction of the population not under stress and neglecting the alarm stage, we assume that susceptible individuals (S) can be split in three subgroups by the types of behavior: *ignorant* or unaware of the epidemic ($S_{ign}$), rationally *resistant* ($S_{res}$), and *exhausted* ($S_{exh}$) that do not react on the external stimuli (this is a sort of refractory period). In other words: $S(t) = S_{ign}(t) + S_{res}(t) + S_{exh}(t)$.

In accordance with this add-on, three additional probabilistic transitions caused by social stress factor emerge in the detailed $SIR_{SS}$ model:

1) $S_{ign} + 2I \rightarrow S_{res} + 2I$ – mobilization reaction (fast). The autocatalytic form here means that the transition rate is proportional to the square of the infected fraction I. This assumption is needed to capture a super-linear increase in alarm reaction to an increase in the number of infected people;

2) $S_{res} \rightarrow S_{exh}$ – exhaustion process due to COVID-19 restrictions (slow);

3) $S_{exh} \rightarrow S_{ign}$ – slow relaxation to the initial state (end of the refractory period).

The first transition represents a stress response to a growing proportion of the infected. The second transition describes the exhaustion of resisting behavior. The third transition completes the loop within the S state and reflects the slow leak of the exhausted people to the state of ignoring the epidemic but susceptible to the alarm signals.

The infection rates of $S_{exh}$ and $S_{ign}$ are assumed to be the same as in the basic SIR model. For the fraction $S_{res}$, the transition to I is much slower, and we take this category as not susceptible to disease.

Possible transitions in the model are illustrated in **Fig. 6**.

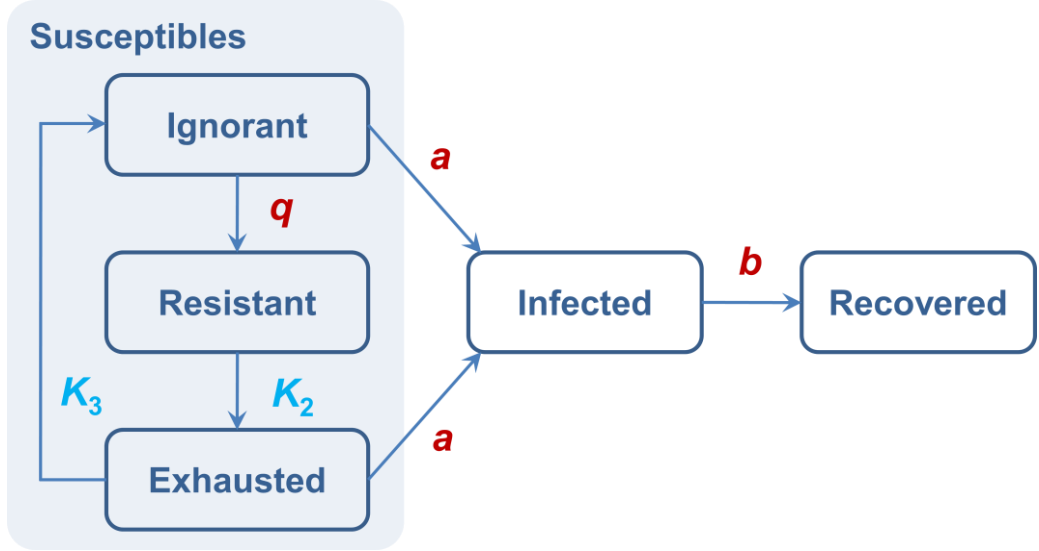

**Fig. 6 A schematic representation of possible transitions in the SIR$_{SS}$ model.** There is a three-stage loop in the S category. Each arrow is associated with a kinetic constant. The rates of transitions "Ignorant → Infected" and "Exhausted → Infected" are also proportional to the infected fraction I, and the rate of transition "Ignorant → Resistant" is proportional to I² (see the autocatalytic representation and equations (1)-(5)). Fast transitions are highlighted in red, and slow transitions are highlighted in light blue.

**Differential equations.** According to the described transitions, the SIR$_{SS}$ model has five variables: $S_{ign}(t)$, $S_{res}(t)$, $S_{exh}(t)$, $I(t)$, $R(t)$, and can be formulated as follows:

$$\frac{dS_{ign}}{dt} = -qS_{ign}I^2 - aS_{ign}I + K_3 S_{exh} \qquad (1)$$

$$\frac{dS_{res}}{dt} = qS_{ign}I^2 - K_2 S_{res} \qquad (2)$$

$$\frac{dS_{exh}}{dt} = -aS_{exh}I + K_2 S_{res} - K_3 S_{exh} \qquad (3)$$

$$\frac{dI}{dt} = aS_{exh}I + aS_{ign}I - bI \qquad (4)$$

$$\frac{dR}{dt} = bI \qquad (5)$$

where $K_1 = q$ is the rate of transition to the resistant state mediated by a stress response in society, $K_2$ is the exhaustion rate due to restrictions ($K_2 \ll 1$), $K_3$ is the constant of transition to a state of ignorance ($K_3 \ll 1$), $K_4 = a$ is the morbidity rate, $K_5 = b$ is the recovery constant.

We kept the following parameters fixed: $b = 0.1$ (recovery rate I → R, expected period of being infected is $b^{-1} = 10$ days), $K_3 = 0.01$ ($S_{exh}$ → $S_{ign}$, expected period of being exhausted is $K_3^{-1}$ = 100 days). Three other parameters: $a$ (the morbidity rate, S + I → 2I), $K_2$ (the exhaustion rate, $S_{res}$ → $S_{exh}$), and $q$ (the stress response rate, $S_{ign}$ + 2I → $S_{res}$ + 2I) were varied to fit experimental data. An additional parameter that affects the rate of the outbreak and, hence, the time of the first wave maximum, is the initial fraction of infected people in the population I(0) = $I_0$ << 1. The rest of the population at the initial state is assumed to be ignorant by default: $S_{ign}$(0) = 1 − $I_0$ ≈ 1.

The choice of parameters $b$ and $K_3$ as constants is determined by the following considerations. The dynamics of the outbreak onset (when $S_{res}$ and $S_{exh}$ are small) depends on the $a/b$ ratio. At large time scale, the trajectory of the epidemic's spread depends on the ratio between $K_2$ and $K_3$. Thus, we choose $b$ and $K_3$ as the references for $a$ and $K_2$, respectively. This means that dynamics is much more sensitive to the $a/b$ and $K_3/K_2$ ratios than to the absolute values of these constants. If we also vary $b$ and $K_3$, we get an unstable (strongly fluctuating in response to a small noise) solution of the optimization problem. Trial simulations showed that the model matches the infection data of COVID-19 for modified values of $b$ and $K_3$ at the same top level preserving $a/b$ and $K_3/K_2$. This matching confirms our choice.

The transition $S_{exh}$ → $S_{ign}$ closes the loop $S_{ign}$ → $S_{res}$ → $S_{exh}$ → $S_{ign}$. It describes the slow transition to the initial *ignorant* state. Qualitatively, this means that humans cannot be exhausted indefinitely. They simply do not respond to external stimuli during a certain refractory period. The transition $S_{exh}$ → $S_{ign}$ significantly affects the dynamics of epidemics. For example, if we take $K_3$ = 0, then the fraction $S_{res}$ will decay faster, and the second wave of the epidemic will quickly absorb most of the population similarly to the first wave in the SIR model (see the right panel in **Fig. 7**). This is contrary to the observed COVID-19 data. **Fig. 7** illustrates the difference between the SIR$_{SS}$ model with and without such a transition.

For this example, we took the SIR$_{SS}$ model fitted to the German data (shown in **Fig. 2**) and compared it with the same model with $K_3$ = 0 and other unchanged reaction rate constants.

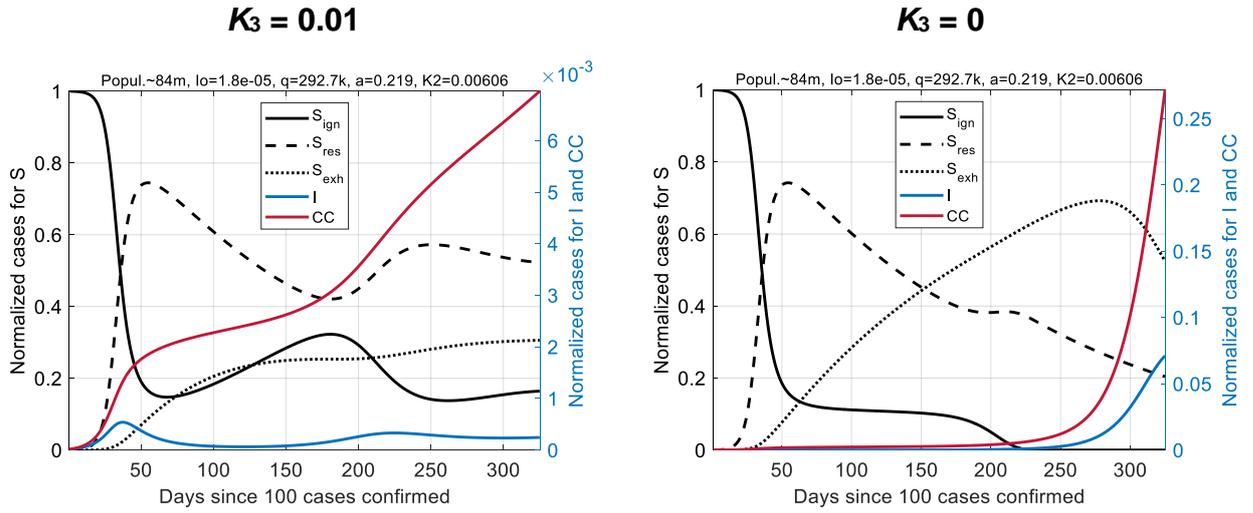

**Fig. 7 Comparison of coronavirus outbreaks at $K_3 = 0.01$ (left panel) and $K_3 = 0$ (right panel): a case for Germany.** The proportion of infected, I, and the number of cases, CC, on the right panel became orders of magnitude higher over time than on the left. Note that the scale for I and CC on the right differs from the left by a factor of 40 (approximately). The second wave on the right will involve almost the entire population.

COVID-19 retrospective data are collected daily. Therefore, all calculations were performed in steps of one day or in steps of an integer number of times less (Δt = 1, 0.5, 0.1, 0.02 etc.). Equations (1)-(5) were transformed into an appropriate discrete-time system.

**Comparison with COVID-19 observations.** Cumulatively confirmed cases of COVID-19 and daily new cases were obtained from a publicly available source of statistical information: the COVID-19 Data Repository by the Center for Systems Science and Engineering (CSSE) at Johns Hopkins University (JHU). The calculation of the dynamics of phase variables according to equations (1)-(5) was carried out using a specially developed MatLab software. The correspondence of the calculated the normalized number of cumulative cases as a function of time, CC(t) = I(t) + R(t), to the normalized observed TCC(t) was estimated using the coefficient of determination $R^2$ (normalization here means dividing by the total population). As the starting point for the algorithm to work (t = 0), we choose the day since at least 100 cases of COVID-19 were confirmed. The time interval for comparing the two samples was equal to the number of days of the modeled trace of CC(t). The search for the global maximum of the $R^2$ was implemented in the parameter space ($a$, $K_2$, $q$, $I_0$), at which the model reproduces epidemics dynamics that are closest to the observed one. The level of $R^2$ is high: for all countries $R^2 > 0.988$, for almost all countries $R^2 > 0.99$, and for seven countries $R^2 > 0.998$. This means that the time dependence TCC(t) is very well described by the trajectories of the $SIR_{SS}$ model on the selected time intervals. The number of parameters to fit is small (four), therefore there is no sign of overfitting.

## Data availability

The series of COVID-19 confirmed cases for all countries of the world are publicly available at data repository of *Our World in Data* project: https://github.com/owid/covid-19-data/tree/master/public/data (direct link to Excel file: https://covid.ourworldindata.org/data/owid-covid-data.xlsx). Raw data come from the COVID-19 Data Repository by the Center for Systems Science and Engineering (CSSE) at Johns Hopkins University (JHU) at https://github.com/CSSEGISandData/COVID-19.

## Code availability

Code used to produce the results presented herein is available in a public GitHub repository at https://github.com/forester52/COVID-19-outbreak-model-with-social-stress-dynamics[27].

## Acknowledgements

The reported study was supported by the Ministry of Science and Higher Education of the Russian Federation, project No. 075-15-2021-634 (conceptual work on the development of a COVID model), and by RFBR according to the research project No. 20-04-60078 (development of computational algorithms, software for data analysis, and simulations).

## Author contributions

I.A.K. and A.N.G. designed the research. I.A.K., E.V.P., and V.B.K. simulated the model. I.A.K. performed data analysis. I.A.K., V.B.K., and A.N.G. interpreted the results. E.M.M. and A.N.G. formulated the model. All authors participated in writing and editing the manuscript.

## Additional information

Correspondence and requests for materials should be addressed to I.A.K.

## Competing interests

The authors declare no competing interests.